\documentclass[%
reprint,
 amsmath,amssymb,
 aps,
]{revtex4-1}

\usepackage{amsmath,amssymb,bm,graphicx}

\newcommand{\bq}{\begin{equation}}
\newcommand{\eq}{\end{equation}}
\newcommand{\bqa}{\begin{eqnarray}}
\newcommand{\eqa}{\end{eqnarray}}
\newcommand{\nn}{\nonumber \\}

\def\be     {\begin{equation}}
\def\ee     {\end{equation}}
\def\bea        {\begin{eqnarray}}
\def\eea        {\end{eqnarray}}
\def\bnn    {\begin{eqnarray*}}
\def\enn    {\end{eqnarray*}}
\def\beqns  {\begin{subequations}\begin{eqnarray}}
\def\eeqns  {\end{eqnarray}\end{subequations}}

\def\REE {{\mathrm{Re} }}
\def\IMM {{\mathrm{Im} }}

\begin{document}

\title{Boltzmann transport theory for many body localization}
\author{ Jae-Ho Han$^{1,2}$ and Ki-Seok Kim$^{1}$ }
\affiliation{$^{1}$Department of Physics, POSTECH, Pohang, Gyeongbuk 790-784, Korea \\ $^{2}$Asia Pacific Center for Theoretical Physics (APCTP), POSTECH, Pohang, Gyeongbuk 790-784, Korea}
\date{\today}

\begin{abstract}
We investigate a many-body localization transition based on a Boltzmann transport theory. Introducing weak localization corrections into a Boltzmann equation,
%
%
Hershfield and Ambegaokar re-derived the Wolfle-Vollhardt self-consistent equation for the diffusion coefficient [Phys. Rev. B {\bf 34}, 2147 (1986)].
%
%
We generalize this Boltzmann equation framework, introducing electron-electron interactions into the Hershfield-Ambegaokar Boltzmann transport theory based on the study of Zala-Narozhny-Aleiner [Phys. Rev. B {\bf 64}, 214204 (2001)]. Here, not only Altshuler-Aronov corrections but also dephasing effects are taken into account. As a result, we obtain a self-consistent equation for the diffusion coefficient in terms of the disorder strength and temperature, which extends the Wolfle-Vollhardt self-consistent equation in the presence of electron correlations. Solving our self-consistent equation numerically, we find a many-body localization insulator-metal transition, where a metallic phase appears from dephasing effects dominantly instead of renormalization effects at high temperatures. Although this mechanism is consistent with that of recent seminal papers [Ann. Phys. (N. Y). {\bf 321}, 1126 (2006); Phys. Rev. Lett. {\bf 95}, 206603 (2005)], we find that our three-dimensional metal-insulator transition belongs to the first order transition, which differs from the Anderson metal-insulator transition described by the Wolfle-Vollhardt self-consistent theory. We speculate that a bimodal distribution function for the diffusion coefficient is responsible for this first order phase transition.
\end{abstract}


\maketitle

\section{Introduction}

Classical particles show their diffusive dynamics in the presence of randomness \cite{Diffusion_Einstein}. On the other hand, quantum interferences give rise to memory effects in the dynamics of quantum particles, which can result in Anderson localization \cite{Anderson_Localization_Anderson}. Various types of inelastic scattering are responsible for dephasing effects, expected to destroy the Anderson localization. Even if electrons of the whole band are localized at zero temperature, electron-phonon interactions erase the memory at an ``infinitesimal" temperature. This transport mechanism is referred to as Mott's variable range hopping \cite{Variable_Range_Hopping_Mott}. What happens in the Anderson localized insulating phase if other sources of delocalized excitations with continuum spectra are neglected and only electron correlations are considered?
%
%

Recently, this fundamental question has been addressed rather rigorously \cite{Many_Body_Localization_Altshuler,Many_Body_Localization_Mirlin}. These studies found that many body wave functions are localized in the functional space below a critical temperature. In other words, a many-body localized insulating state is still stable up to a critical temperature, where the electrical conductivity remains to be zero identically. Taking into account an exponentially decaying localized one-particle wave function, level repulsion for the states localized nearby, and randomly signed matrix elements of the interaction decay
%
%
\cite{Introduction_MBL_Altshuler}, Ref. \cite{Many_Body_Localization_Altshuler} showed that the probability distribution function $P(\Gamma)$ for the imaginary part $\Gamma$ of the self-energy is given by $P(\Gamma) \sim \delta(\Gamma)$ in the many-body localized insulating phase below the critical temperature and $P(\Gamma > 0) \not= 0$ in the metallic state above the critical temperature.

Since these seminal papers, the research on many-body localization has been performed extensively \cite{MBL_1,MBL_2,MBL_3,MBL_4,MBL_5,MBL_6,MBL_7,MBL_8,MBL_9,MBL_10,MBL_11,MBL_12,MBL_13,MBL_14,
MBL_15,MBL_16,MBL_17,MBL_18,MBL_19,MBL_20,MBL_21,MBL_22,MBL_23,MBL_24,MBL_25,MBL_26,MBL_27,MBL_28,MBL_29,MBL_30,MBL_31,MBL_32,MBL_33}. They have discussed the space-time evolution of entanglement properties, the nature of level statistics and eigenfunction correlations, properties of distributions in coupling functions, anomalous diffusive transport phenomena, and more fundamental understanding in connection with an infinite randomness fixed point and its involved Griffiths phase, integrability, and etc. However, there do not exist any studies which show the emergence of a many-body localized insulating state from a metallic phase, taking into account resumming contributions of electron interactions, i.e., Feynman diagrams in a diffusive metallic phase, as far as we know. In the present study we investigate a many-body localization transition based on the Boltzmann transport theoretical framework.

Introducing the memory effect from weak localization into a Boltzmann equation, Hershfield and Ambegaokar re-derived the Wolfle-Vollhardt self-consistent equation \cite{Self_Consistent_Theory_AMIT} for the diffusion coefficient \cite{Boltzmann_Transport_Theory_AMIT}.
%
%
In this study we extend this Boltzmann equation framework, introducing electron-electron interactions into the Hershfield-Ambegaokar Boltzmann transport theory based on the study of Zala-Narozhny-Aleiner \cite{Interaction_Boltzmann_Transport_Theory}. Here, not only Altshuler-Aronov corrections \cite{Altshuler_Aronov} but also dephasing effects \cite{Dephasing} are taken into account in the presence of weak localization corrections.
%
%
%
Based on this generalized Boltzmann transport theory, we find that our three-dimensional metal-insulator transition shows the first order transition, which differs from the Anderson metal-insulator transition described by the Wolfle-Vollhardt self-consistent theory. We suspect that a bimodal distribution function for the diffusion coefficient is responsible for this first order phase transition.

Before going further, we would like to point out a recent study, which examined a many-body localization transition in two dimensions from a ``metallic" side \cite{MBL_AAK_RG}. Based on the Altshuler-Aronov-Khmelnitsky path integral formulation for the Cooperon propagator \cite{Dephasing}, this study proposed a second quantized representation. Resorting to the replica trick, this field theory formulation reproduced the dephasing rate of the previous diagrammatic studies in the self-consistent Born approximation. In order to describe a metal-insulator transition, these authors performed the perturbative renormalization group analysis beyond the mean-field theory. They found an unstable fixed point characterized by the fact that the dephasing rate vanishes, referred to as dephasing catastrophe. They speculated that this fixed point of dephasing catastrophe may describe a many-body localization transition in two dimensions. As well discussed in this study, these authors did not take into account a scale dependent self-consistency condition between the renormalization group controlling the dephasing due to real processes and the running renormalization of the diffusion constant due to the virtual ones \cite{Dephasing}. On the other hand, we suggest the self-consistency condition for the diffusion constant as a function of both the disorder strength and temperature.

\section{From weak localization to Anderson metal-insulator transition in the Boltzmann equation approach}

In this section we review the Hershfield-Ambegaokar Boltzmann transport theory to show how the Wolfle-Vollhardt self-consistent equation is derived to describe an Anderson metal-insulator transition in a continuous manner. Although there are some results not reported before as far as we know, readers who are familiar to this self-consistent theory may skip the present review section.

\subsection{Boltzmann transport theory with weak localization}

We start from a Boltzmann equation to describe the evolution of a distribution function $f(t, \bm r, \bm p)$, given by
\begin{eqnarray}
\bigg( \frac{\partial}{\partial t} + \dot{\bm r} \cdot \bm \nabla_r + \dot{\bm p} \cdot \bm \nabla_p \bigg) f(t, \bm r, \bm p) = \bigg( \frac{\partial f}{\partial t} \bigg)_{coll} . \label{Boltzmann_Eq}
\end{eqnarray}
One can derive the Boltzmann equation from the Schrodinger equation, based on the Schwinger-Keldysh formulation, where the distribution function is given by the lesser Green's function \cite{Mahan_Textbook}. In this formulation $t$ and ${\bm r}$ denote the center-of-mass time and space coordinates of a particle-hole pair, respectively. ${\bm p}$ represents an internal momentum between the particle-hole pair, the Fourier-transformed variable of a relative coordinate. Here, the internal frequency of $\omega$ is integrated over to give a classical Boltzmann equation. Then, ${\bm r}$ and ${\bm p}$ may be regarded to be coordinates in the phase space.
%
%
$\dot{\bm r} = \bm v_F$ is the group velocity, here the Fermi velocity, and $\dot{\bm p} = e\bm E$ is nothing but the classical equation of motion.

The collision term describes elastic scattering between a quasiparticle and nonmagnetic impurities, given by
\begin{eqnarray}
&&\bigg( \frac{\partial f}{\partial t} \bigg)_{coll}
= -\frac{f(t, \bm r, \bm p) - f_{eq}(\bm p)}{\tau} \nn
&&\hspace{30pt} + \int_{-\infty}^t\!dt' \ \alpha(t-t') \Big[ f(t', \bm r, -\bm p) - f_{eq}(\bm p) \Big] .
\end{eqnarray}
The first term is obtained in the relaxation-time approximation, where $\tau$ is the mean-free time to describe the average time between collisions \cite{Kardar_Textbook}. This collision term follows from the self-energy correction in independent Born-approximation scattering from impurities \cite{Boltzmann_Transport_Theory_AMIT}, responsible for the diffusive motion of the quasiparticle. The last term describes quantum mechanical interference effects of the quasiparticle wavefunction, given by a non-local term in time and involved with multiple elastic scattering. Actually, the maximally crossed diagram gives rise to the nonlocal lesser self-energy in terms of the memory kernel $\alpha(t-t')$ and the lesser Green's function (distribution function) $f(t', \bm r, -\bm p) - f_{eq}(\bm p)$, where the memory kernel is nothing but the diffusion propagator \cite{Boltzmann_Transport_Theory_AMIT}. Phase accumulation due to such multiple disorder scattering gives rise to weak localization \cite{Altshuler_Aronov}. It is interesting to observe $- {\bm p}$ in $f(t', \bm r, -\bm p)$, implying a time-reversal path of a quasiparticle.

The weak-localization memory kernel is given by \cite{Boltzmann_Transport_Theory_AMIT}
\begin{eqnarray}
\alpha (t-t') = 2n_i |V|^2 \int'\!\frac{d^3q}{(2\pi)^3} \ e^{- \left( D_{0} q^2 + \frac{1}{\tau_\phi} \right) (t-t')} ,
\end{eqnarray}
which corresponds to the momentum integral of the diffusion propagator in the frequency space. $n_{i}$ is the density of nonmagnetic impurities and $V$ is the potential strength of disorder. More precisely, $|V|^2$ is the variance of the disorder potential. In the kernel $D_{0} = v_F l /d$ is the diffusion coefficient of a quasiparticle in $d$ spatial dimension, where $l=v_F\tau$ is the mean-free path, and $\tau_{\phi}$ is the maintenance time of phase coherence. The $q$-integral $\int'$ has the upper and lower cutoffs, given by $1/\tau_\phi < |\bm q| < 1/l$. Here, the inverse of the coherence time plays the role of the lower cutoff. The scattering rate $\Gamma = 1/\tau$ can be found in the Born approximation, given by $\Gamma = 2 \pi n_i |V|^2 N_F$, where $N_{F}$ is the density of states at the Fermi energy \cite{Altshuler_Aronov}.

It is straightforward to solve this Boltzmann equation under a uniform electric field $\bm E$. Assuming homogeneity of our system, we have $f(t, \bm r, \bm p) \rightarrow f(t, \bm p)$. Resorting to the spherical coordinate for the momentum space, we have $f(t, \bm p) \rightarrow f(t, \bm n, \epsilon)$, where $\bm n = \bm p / |\bm p|$ is an angular direction and $\epsilon = p^2/2m$ is the dispersion relation of a quasiparticle. Then, we introduce the following ansatz for the solution of the Boltzmann equation
\begin{eqnarray}
f(t, \bm n, \epsilon) = f_{eq}(\epsilon) + \bm n \cdot \bm \Gamma(t, \epsilon) . \label{Solution_Ansatz}
\end{eqnarray}
$f_{eq}(\epsilon)$ is an equilibrium distribution function to describe an isotropic system. $\bm n \cdot \bm \Gamma(t, \epsilon)$ describes the variation of the distribution function up to the linear order of the applied electric field.

Inserting this ansatz into the Boltzmann equation, we obtain
\begin{eqnarray}
\frac{\partial}{\partial t} \Gamma_\alpha (t, \epsilon) + ev_F E_\alpha(t) \ \partial_\epsilon f_{eq}(\epsilon) = -\frac{\Gamma_\alpha(t, \epsilon)}{\tau} \nn
- \int_{-\infty}^t\!dt' \ \alpha(t-t') \Gamma_\alpha(t', \epsilon) ,
\end{eqnarray}
where the subscript $\alpha = 1, ..., d$ is the direction of momentum in $d$ dimensions. In order to solve the memory effect, we perform the Fourier transformation and obtain
\begin{eqnarray}
&&-i\omega \Gamma_\alpha (\omega, \epsilon) + ev_F E_\alpha(\omega) \ \partial_\epsilon f_{eq}(\epsilon) \nn
&&\hspace{80pt}= -\frac{\Gamma_\alpha(\omega, \epsilon)}{\tau} - \alpha(\omega) \Gamma_\alpha(\omega, \epsilon) .
\end{eqnarray}
Here, the memory kernel is given by
\begin{eqnarray}
\alpha(\omega)\tau
&=& \frac{1}{\pi N_F} \int'\!\frac{d^3q}{(2\pi)^3} \ \frac{1}{-i\omega + D_{0} q^2 + \frac{1}{\tau_\phi}} \nn
&=& \pi \lambda^2 \int_0^1\!d\tilde q \frac{\tilde q^2}{-i\omega\tau + {1\over3} \tilde q^2},
\end{eqnarray}
which is nothing but the momentum integral of the diffusion propagator. In the second equality, we introduced dimensionless quantities of $\tilde q = ql$ and $\lambda = 1/(\pi k_F l)$. The latter describes disorder strength, where $\lambda = 0$ corresponds to the clean case. Also, we take $\tau_\phi/\tau \rightarrow \infty$ in the second equation, safely allowed in $d=3$. Note that $\alpha(\omega)\tau$ is proportional to $\lambda^2$. Then, the shift of the distribution function is
\begin{eqnarray}
\Gamma_\alpha(\omega, \epsilon) = -\frac{ev_F \partial_\epsilon f_{eq}(\epsilon)}{-i\omega + 1/\tau + \alpha(\omega)} E_\alpha(\omega).
\end{eqnarray}

Introducing the distribution function into the formal expression of the electrical current, we obtain
\begin{eqnarray}
\bm j(\omega)
&=& e\int\!\frac{d^3p}{(2\pi)^3} \ \bm v_p f(\omega, \bm p) \nn
&=& \frac{N e^2 v_F^2}{-i\omega + 1/\tau + \alpha(\omega)} \left< \bm n \big( \bm n \cdot \bm E(\omega) \big) \right>_n \nn
&=& \frac{N e^2 v_F^2/3}{-i\omega + 1/\tau + \alpha(\omega)} \bm E(\omega) ,
\end{eqnarray}
where $\left< ... \right>_n$ denotes an integral for the angle average and the numerical factor $1/3$ results from $d = 3$. Then, the electrical conductivity reads
\begin{eqnarray}
\frac{\sigma(\omega)}{\sigma_0}
&=& \frac{1}{1-i\omega\tau + \alpha(\omega)\tau} \nn
&=& \left( 1 - i\omega\tau + \frac{1}{\pi N_F} \int'\!\frac{d^3q}{(2\pi)^3} \ \frac{1}{-i\omega + D_{0} q^2 + \frac{1}{\tau_\phi}} \right)^{-1} \nn
&\approx& \frac{1}{ 1 - i\omega\tau } \nn
&&- \frac{1}{ (1 - i\omega\tau)^{2} }\frac{1}{\pi N_F} \!\!\int'\!\!\frac{d^3q}{(2\pi)^3} \ \frac{1}{-i\omega + D_{0} q^2 + \frac{1}{\tau_\phi}}, \label{RPA_WL}
\end{eqnarray}
where $\sigma_0 = \frac{1}{3}N_F e^2 v_F^2\tau$ is the dc Drude conductivity. The second line shows the well-known expression for the weak-localization correction in the limit of $\lambda \ll 1$.

\subsection{Wolfle-Vollhardt self-consistent theory for an Anderson metal-insulator transition}

\begin{figure}[t!]
\includegraphics[width=8.5cm]{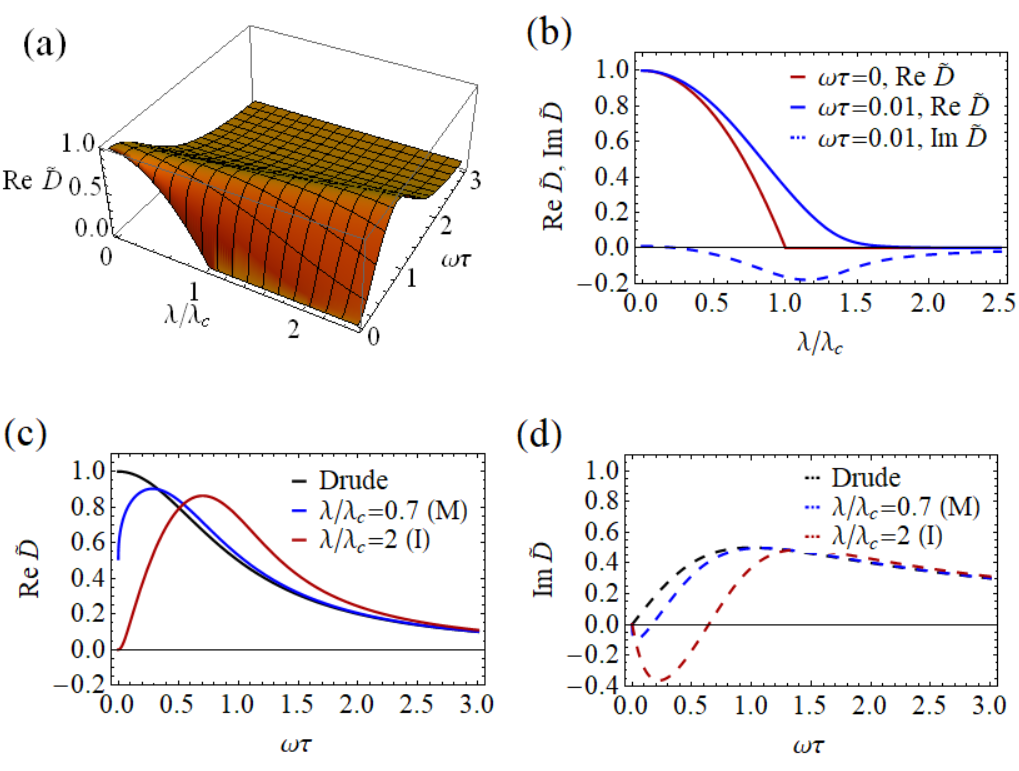}
\caption{Diffusion coefficient as a function of the disorder strength $\lambda = 1 / (\pi k_{F} l)$ and the external frequency $\omega \tau$ in the W\"olfle-Vollhardt self-consistent theory for an Anderson metal-insulator transition. Here, $k_{F}$ is the Fermi momentum, $l$ ($\tau$) is the mean-free path (time), and $\lambda_c = 1/\sqrt{3\pi}$. (a) shows the diffusion coefficient as a function of disorder strength $\lambda/\lambda_c$ and external frequency $\omega\tau$.  (b) shows the diffusion coefficient as a function of the disorder strength, where the red-thick line denotes the dc diffusion coefficient and the blue-thick (dashed) line does the real (imaginary) part of the ac diffusion coefficient at $\omega \tau = 0.01$. (c) displays the real part of the ac diffusion coefficient as a function the external frequency, where the blue-thick (red-thick) line denotes it in the metallic, denoted by $M$,  (insulating, denoted by $I$) phase of $\lambda/\lambda_c = 0.7$ ($\lambda/\lambda_c = 2$). (d) exhibits the imaginary part of the ac diffusion coefficient as a function the external frequency, where the blue-thick dashed (red-dashed) line expresses it in the metallic (insulating) phase of $\lambda/\lambda_c=0.7$ ($\lambda/\lambda_c = 2$). For the comparison, we also show the Drude conductivity as the black-thick (black-dashed) line.}
\label{fig01}
\end{figure}

\begin{figure*}[t!]
\includegraphics[width=18cm]{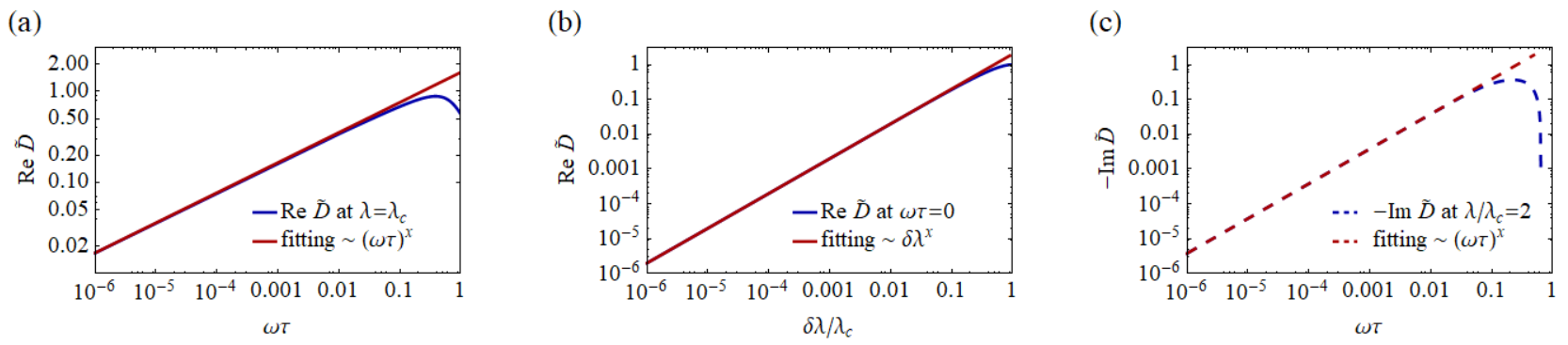}
\caption{The power-law behavior near the transition point, $\lambda=\lambda_c = 1/\sqrt{3\pi}$ with $\omega\tau=0$. (a) The blue line is a log-log plot of $\REE\tilde D$ versus $\omega\tau$ with $\lambda=\lambda_c$. The red line is a fitting curve of the form $(\omega\tau)^x$, where $x\approx 0.330$. (b) The blue line is a log-log plot of $\REE\tilde D$ versus $\delta\lambda/\lambda_c$ with $\omega\tau=0$. The red line is a fitting curve of the form $\delta\lambda^x$ with $x\approx 0.998$. (c) The blue-dashed line is a log-log plot of $-\IMM\tilde D$ versus $\omega\tau$ with $\lambda/\lambda_c=2$ which is in the insulating phase. The red-dashed line is a fitting curve of the form $\delta\lambda^x$ with $x\approx 1.000$.}
\label{fig02}
\end{figure*}

It is natural to consider that an Anderson metal-insulator transition may be understood by some types of resummations for weak localization corrections. We recall Eq. (\ref{RPA_WL}) with the Einstein relation, given by
\begin{eqnarray}
\frac{D(\omega)}{D_0}
= \Bigg( 1 - i\omega\tau + \frac{1}{\pi N_F} \!\!\int'\!\!\frac{d^3q}{(2\pi)^3} \ \frac{1}{-i\omega + D_{0} q^2 + \frac{1}{\tau_\phi}} \Bigg)^{-1}. \nonumber
\end{eqnarray}
An idea is to replace the bare value of the diffusion coefficient $D_{0}$ in the weak-localization memory kernel with a fully renormalized one as follows \cite{Boltzmann_Transport_Theory_AMIT}
\begin{eqnarray}
\frac{D(\omega)}{D_0} = \left( 1 - i\omega\tau + \frac{1}{\pi N_F} \!\!\int'\!\!\frac{d^3q}{(2\pi)^3} \frac{1}{-i\omega + D(\omega) q^2 + \frac{1}{\tau_\phi}} \right)^{-1} \nn
\end{eqnarray}
This is the self-consistent equation of Wolfle-Vollhardt theory for an Anderson metal-insulator transition \cite{Self_Consistent_Theory_AMIT}. Introducing dimensionless parameters of $\tilde D(\omega) = D(\omega)/D_0$, $\tilde q = ql$, and $\lambda = 1/(\pi k_Fl)$, we can rewrite the above equation as
\begin{eqnarray}
\tilde D(\omega) = \left( 1 - i\omega\tau + \pi \lambda^2 \int_0^1\!d\tilde q \ \frac{\tilde q^2}{-i\omega\tau + {1\over3}\tilde D(\omega) \tilde q^2} \right)^{-1}. \nn
\label{eq:SC}
\end{eqnarray}

We solve this equation numerically and find the real and imaginary parts of the diffusion coefficient as a function of the disorder strength $\lambda$ and the external frequency $\omega \tau$, shown in Fig. 1(a). Increasing the disorder strength, the diffusion coefficient decreases gradually and vanishes at the critical strength of disorder $\lambda = 1/\sqrt{3\pi} \equiv \lambda_c$ in the dc limit, shown in Fig. 1(b). This is a typical signature of the second order phase transition. On the other hand, the signature of a sharp phase transition is smeared in the real part of the diffusion coefficient at a finite frequency. The imaginary part of the diffusion coefficient shows a hump around the critical point. It is interesting to see that the weak localization correction gives rise to a dip of the diffusion coefficient at zero frequency in a metallic phase, where the maximum is shifted into a finite frequency. See Fig. 1(c). The dip becomes deeper as the disorder strength increases. Exceeding the critical disorder strength, the diffusion coefficient vanishes at zero frequency as expected, and the maximum value of the real part is shifted into a larger frequency. The weak localization correction results in the sign change to the imaginary part of the diffusion coefficient, where the frequency of the sign change increases as the disorder strength is enhanced, shown in Fig. 1(d).

\subsection{Scaling theory for the diffusion coefficient}

The above numerical solution can be more quantified, based on the scaling theory given by
\begin{eqnarray}
D(\delta\lambda, \omega)
&=& b^{-y_D} D\big( b^{y_{\delta\lambda}} \delta\lambda,  b \omega \big) \nn
&=& \left\{ \begin{array}{ll}
\omega^{y_D} D \left( \frac{\delta\lambda}{\omega^{y_{\delta\lambda}}}, 1 \right), & b = \omega^{-1}, \\
\delta\lambda^{y_D \over y_\lambda}  D \left( 1, \frac{\omega}{\delta\lambda^{1/ y_{\delta\lambda}}} \right), & b = \delta\lambda^{-{1 \over y_{\delta\lambda}}},
\end{array}\right.
\end{eqnarray}
where $\delta\lambda = \lambda_c - \lambda$. First, we set $\delta\lambda=0$. The log-log plot of $\REE \tilde D$ as a function of $\omega\tau$ is shown in Fig. 2(a) (blue curve). It shows that $\REE\tilde D$ has a power-law behavior as $\omega\tau$ goes to the critical value $\omega\tau=0$. Fitting the curve with a function of the form proportional to $(\omega\tau)^x$ on the interval $0<\omega\tau < 10^{-4}$, we obtain $x\approx 0.330$, which is very close to $1/3$. Second, we set $\omega\tau=0$. The log-log plot of $\REE\tilde D$ as a function of $\delta\lambda$ is shown in Fig. 2(b) (blue curve). It shows that $\REE\tilde D$ has a power-law behavior as $\delta\lambda$ goes to the critical value $\delta\lambda=0$. Fitting the curve with a function of the form proportional to $\delta\lambda^x$ on the interval $0<\delta\lambda/\lambda_c<10^{-4}$, we obtain $x\approx 0.998$ which is very close to $1$. These critical exponents suggest $y_D\approx 1/3$ and $y_{\delta\lambda}\approx 1/3$. In Fig. \ref{fig02} (c), we plot $-\IMM\tilde D$ versus $\omega\tau$ (blue-dashed line) in the insulating phase ($\lambda/\lambda_c=2$), where the fitting curve (red-dashed line) suggests the form $(\omega\tau)^x$ with $x$ as a fitting parameter on the interval $0<\omega\tau<10^{-4}$. We obtain $x\approx 1.000$ from this fitting. This implies that the localization length $\xi$ defined by
\begin{eqnarray}
\xi = \lim_{\omega\rightarrow 0} \sqrt{\frac{D(\omega)}{-i\omega}}
\end{eqnarray}
is finite in the insulating phase ($\REE D = 0$), while it is divergent in the metallic phase ($\REE D \neq 0$).

Some critical exponents can be obtained analytically from the self-consistent equation, Eq. (\ref{eq:SC}). As shown in the above, $\xi \rightarrow \infty$ in the metallic phase, so from Eq. (\ref{eq:SC}) we have
\begin{eqnarray}
\tilde D(\omega=0) = 1-\frac{\lambda^2}{\lambda_c^2} \propto |\lambda_c-\lambda|^s, \ \ \ s= {y_D \over y_{\delta\lambda}} = 1.
\end{eqnarray}
In the insulating phase given by $\REE D(\omega=0) = \IMM D(\omega=0)=0$, where $\xi$ is finite, Eq. (\ref{eq:SC}) implies that \cite{Anomalous_Diffusion}
\begin{eqnarray}
\xi \propto |\lambda-\lambda_c|^{-\nu}, \ \ \ \nu =1.
\end{eqnarray}
%
%

\begin{figure}[t!]
\includegraphics[width=8.5cm]{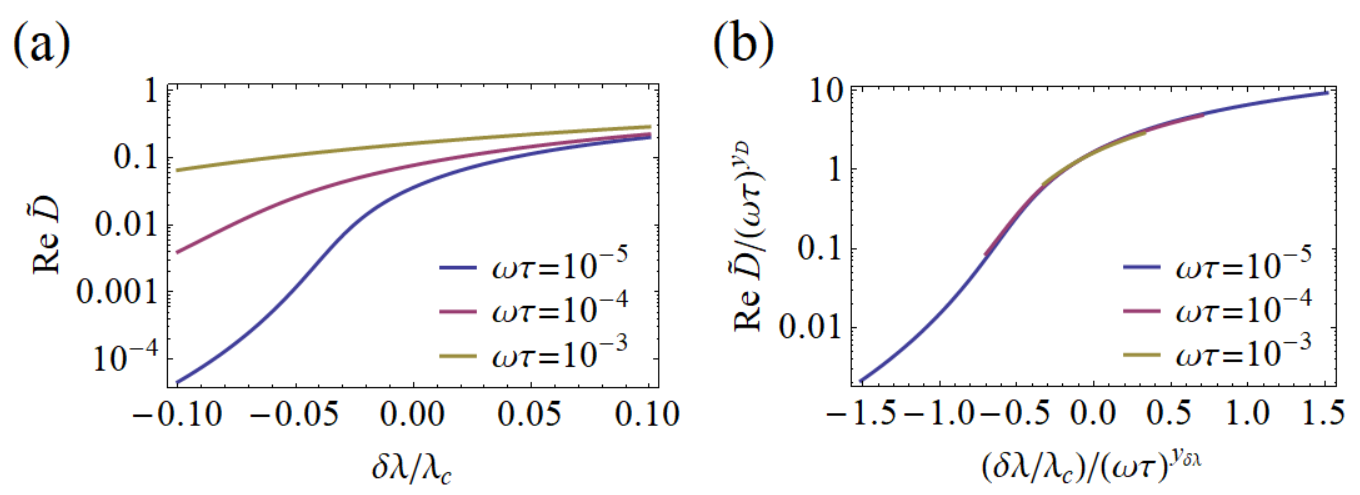}
\caption{(a) Log-linear plot of $\REE \tilde D$ versus $\delta\lambda/\lambda_c$ with three different values of $\omega\tau$, as shown in the inset. (b) Log-linear plot of $\REE \tilde D/(\omega\tau)^{y_D}$ versus $(\delta\lambda/\lambda_c)/(\omega\tau)^{y_{\delta\lambda}}$.}
\label{fig03}
\end{figure}

Figure 3(a) shows $\REE\tilde D$ as a function of $\delta\lambda/\lambda_c$ with three different values of $\omega\tau$ close to the critical value $\omega\tau=0$. Considering the rescaling of $\REE\tilde D \rightarrow \REE\tilde D/(\omega\tau)^{y_D}$ and $\delta\lambda/\lambda_c \rightarrow (\delta\lambda/\lambda_c) /(\omega\tau)^{y_{\delta\lambda}}$ with the critical exponents above, the three curves collapse into a single curve, as the scaling theory predicts. See Fig. 3(b).

\section{Introduction of electron-electron interaction corrections into Boltzmann transport theory with weak localization}

In this section we generalize the Hershfield-Ambegaokar Boltzmann transport theory, introducing electron correlations into the Wolfle-Vollhardt self-consistent equation for the diffusion coefficient based on the study of Zala-Narozhny-Aleiner. Self-consistency will be taken into account in section IV.

\subsection{Electron-electron interaction corrections in the Hershfield-Ambegaokar Boltzmann transport theory}

%
%
Recalling Eq. (\ref{Boltzmann_Eq}), we write down a collision term
\begin{eqnarray}
&&\bigg( \frac{\partial f}{\partial t} \bigg)_{coll, el}
= -\frac{f(t, \bm r, \bm p) - f_{eq}(\bm p)}{\tau} \nn
&&+ \int_{-\infty}^t\!\!dt' \ \alpha(t-t') \Big[ f(t', \bm r, -\bm p) - f_{eq}(\bm p) \Big] + I_{ee} ,
\end{eqnarray}
which takes into account electron-electron interactions, $I_{ee}$. This electron-electron collision term is well evaluated in Ref. \cite{Interaction_Boltzmann_Transport_Theory}. The electron collision term consists of two contributions, given by
\begin{eqnarray}
I_{ee}
&=& I_0(t, \bm r, \epsilon, \bm n) \Big< f(t, \bm r, \epsilon, \bm n) \Big>_n \nn
&&\hspace{30pt}+ n_\alpha I_1^{\alpha\beta}(t, \bm r, \epsilon) \Big< n_\beta f(t, \bm r, \epsilon, \bm n) \Big>_n ,
\end{eqnarray}
where
\beqns
&&I_0(t, \bm r, \epsilon, \bm n)
= -\frac{8}{\tau} \int\!\frac{d\omega}{2\pi} \Bigg\{ n_\alpha K_0^{\alpha\beta}(\omega) \Big< n_\beta f(t, \bm r, \epsilon - \omega, \bm n) \Big>_n \nn
&&\hspace{10pt}+ \frac{n_\alpha L_0^{\alpha\beta}(\omega)}{2} \left(\nabla_\beta + eE_\beta \frac{\partial}{\partial\epsilon} \right) \Big< f(t, \bm r, \epsilon - \omega, \bm n) \Big>_n \Bigg\}, \\
&&I_1^{\alpha\beta}(t, \bm r, \epsilon)
= -\frac{8}{\tau} \int\!\frac{d\omega}{2\pi} \ K_1^{\alpha\beta}(\omega) \Big< f(t, \bm r, \epsilon - \omega, \bm n) \Big>_n.
\eeqns

Interaction kernels are given by \cite{Interaction_Boltzmann_Transport_Theory}
\beqns
K_0^{\alpha\beta}(\omega)
&=& \IMM \int\!\frac{d^3q}{(2\pi)^3} \ \mathcal{D}^R(\omega, \bm q) \bigg\{ \big< n_\alpha D n_\beta \big> \big< D \big> \nn
&&- \frac{i}{v_F} \frac{\partial}{\partial q_\alpha} \big< Dn_\beta \big> - \big< Dn_\alpha \big> \big< Dn_\beta \big> \bigg\}, \\
K_1^{\alpha\beta}(\omega)
&=& \IMM \int\!\frac{d^3q}{(2\pi)^3} \ \mathcal{D}^R(\omega, \bm q) \bigg\{ \big< n_\alpha D \big> \big< Dn_\beta \big> \nn
&&- \frac{\delta_{\alpha\beta}}{2} \left( \big< D \big> \big< D \big> + i\frac{\partial}{\partial\omega} \big< D \big> \right) \bigg\}, \\
L_0^{\alpha\beta}(\omega)
&=& -\REE \int\!\frac{d^3q}{(2\pi)^3} \ \mathcal{D}^R(\omega, \bm q) \bigg\{ \big< D \big> \frac{\partial}{\partial q_\beta} \big< n_\alpha D \big> \nn
&&- \big< Dn_\alpha \big> \frac{\partial}{\partial q_\beta} \big< D \big> - \left< Dn_\alpha \frac{\partial}{\partial q_\beta} D \right> \bigg\} .
\eeqns
They consist of two parts. One is an effective interaction retarded propagator $\mathcal{D}^R(\omega, \bm q)$, where the electrical conductivity is calculated in the lowest order for interaction corrections, referred to as the renormalized Hartree-Fock approximation. This interaction propagator is given by the random phase approximation (RPA) for particle-hole excitations, where the vertex of the particle-hole bubble is renormalized by the diffusion ladder. The other is the electron-hole bubble of the current-current correlation function, where the current vertex is also renormalized by the diffusion ladder. In other words, one RPA renormalized interaction line appears in the particle-hole bubble diagram of the current-current correlation function, where not only the interaction vertex in the Hartree-Fock diagram but also the current vertex is renormalized by the diffusion ladder. Here, $D$ is the propagator to describe the classical motion of a quasiparticle on the Fermi energy, given by the solution of the following equation
\begin{eqnarray}
&&\Big( -i\omega + iv_F \bm n \cdot \bm q \Big) D(\bm n, \bm n'; \omega, \bm q) \nn
&&+ \frac{1}{\tau} \Big[ D(\bm n, \bm n'; \omega, \bm q) - \big< D(\bm n, \bm n'; \omega, \bm q) \big>_n \Big] = c \delta(\bm n, \bm n'), \nn
\end{eqnarray}
where $c=4\pi$ ($c=2\pi$) for $d=3$ ($d = 2$). Integrals for the angle average are shown in Appendix A for three dimensions and Appendix B for two dimensions.

In the absence of external magnetic fields, we have $K_i^{\alpha\beta} = \delta_{\alpha\beta} K_i$ and $L_0^{\alpha\beta} = \delta_{\alpha\beta} L_0$. These interaction kernels are \cite{Interaction_Boltzmann_Transport_Theory}
\beqns
K_0(\omega)
&=& \IMM \ \frac{1}{2}\int\!\frac{d^3q}{(2\pi)^3} \ \mathcal{D}^R(\omega, q) \Bigg\{ \frac{C - \big( -i\omega + 1/\tau \big)}{C \big( C - 1/\tau \big)^2} \nn
&&+ \frac{\big[ C - \big( -i\omega + 1/\tau \big) \big]^2}{C \big( C-1/\tau \big)} \frac{\tau}{3D_0 q^2} \Bigg\}, \\
K_1(\omega)
&=& -\IMM \ \frac{1}{2} \int\!\frac{d^3q}{(2\pi)^3} \ \mathcal{D}^R(\omega, q) \Bigg\{ \frac{C-\big( -i\omega + 1/\tau \big)}{C \big( C-1/\tau \big)^2} \nn
&&+ \frac{\tau}{3D_0q^2} \left( \frac{C-\big( -i\omega + 1/\tau \big)}{C-1/\tau} \right)^2 \Bigg\}, \\
\frac{L_0(\omega)}{v_F\tau}
&=& -\IMM \ \frac{1}{2} \int\!\frac{d^3q}{(2\pi)^3} \ \mathcal{D}^R(\omega, q) \Bigg\{ \frac{9}{2\tau^2} \frac{D_0q^2}{C^3 \big( C-1/\tau \big)^2} \nn
&&+ \frac{3D_0q^2}{C^3} \frac{1/\tau^3}{\big( C-1/\tau \big)^3} \Bigg\},
\eeqns
where
\begin{eqnarray}
\mathcal{D}^R(\omega, q) \approx -\frac{1}{N_F} \frac{-i\omega + {3\over2} D_0 q^2}{{3\over2} D_0 q^2} , \ \ \ C \approx \frac{1}{\tau} - i\omega + {3\over2} D_0 q^2 . \nn \label{Interaction_Propagator}
\end{eqnarray}
Here, we expressed the kernels in terms of $D_0$ by using the relation of $v_F^2 = 3D_0/\tau$.

The collision term of $I_1^{\alpha\beta}(t, \bm r, \epsilon)$ describes electron scattering by Friedel oscillations due to static disorder, given by the Hartree-Fock approximation and referred to as Altshuler-Aronov corrections \cite{Interaction_Boltzmann_Transport_Theory}. On the other hand, $I_0(t, \bm r, \epsilon, \bm n)$ expresses electron scattering by non-equilibrium nonlocal Fock-like potential created by all other electrons \cite{Interaction_Boltzmann_Transport_Theory}. These interactions are real processes, responsible for dephasing, while Altshuler-Aronov corrections are virtual processes. Based on this Boltzmann equation framework, we show that such dephasing processes are responsible for a many-body localization insulator-metal transition at a finite temperature.

Several remarks are in order. First, we did not take into account spin triplet interaction channels, where only spin singlet charge channels are introduced into the Boltzmann equation. Second, the interaction strength does not appear in the interaction retarded propagator as long as only the spin singlet channel is considered. As a result, the electrical conductivity does not depend on the interaction strength. Mathematically speaking, this originates from RPA for the renormalized interaction propagator, given by the inverse of the particle-hole polarization function for the charge channel \cite{Interaction_Boltzmann_Transport_Theory}. Zala-Narozhny-Aleiner claimed that their interaction corrections in the electrical conductivity hold as long as the Landau's Fermi liquid state is preserved, i.e., no symmetry breaking occurs \cite{Interaction_Boltzmann_Transport_Theory}. Third, we focus on the diffusive regime, given by $T \tau \ll 1$, where $T$ is temperature. This diffusive dynamics has been introduced into the effective interaction propagator of Eq. (\ref{Interaction_Propagator}).

\subsection{Electrical conductivity in the presence of both weak localization and electron interaction corrections}

Following the previous section, we use the ansatz $f(t, \bm n, \epsilon) = f_{eq}(\epsilon) + \bm n \cdot \bm \Gamma(t, \epsilon)$ and linearize the Boltzmann equation in $\bm E$. Then, we obtain
\begin{eqnarray}
&&\Gamma_\alpha(\omega, \epsilon) \nn
&\approx& - \frac{e v_F \tau}{1-i\omega\tau + \alpha(\omega) \tau} E_\alpha(\omega) \frac{\partial f_{eq}(\epsilon)}{\partial \epsilon}\nn
&& -\frac{4 f_{eq}(\epsilon)}{1-i\omega\tau + \alpha(\omega) \tau} \int\!\frac{d\epsilon'}{2\pi} \ L_0^{\alpha\beta}(\epsilon') eE_\beta(\omega) \frac{\partial}{\partial\epsilon} f_{eq}(\epsilon - \epsilon') \nn
&&+ \frac{8 e v_F \tau}{3\big[ 1-i\omega\tau + \alpha(\omega) \tau \big]^2} \int\!\frac{d\epsilon'}{2\pi} \ \bigg[ K_1^{\alpha\beta}(\epsilon') f_{eq}(\epsilon - \epsilon')  \nn
&&\hspace{30pt}\times \frac{\partial f_{eq}(\epsilon)}{\partial \epsilon} + K_0^{\alpha\beta} (\epsilon') f_{eq}(\epsilon) \frac{\partial f_{eq}(\epsilon - \epsilon')}{\partial \epsilon} \bigg] E_\beta(\omega) \nn
&& + \frac{32e}{3\big[ 1 + i\omega\tau + \alpha(\omega)\tau \big]^2} \int\!\frac{d\epsilon'd\epsilon''}{(2\pi)^2} \bigg[ K_1^{\alpha\beta}(\epsilon')f_{eq}(\epsilon-\epsilon') \nn
&&\hspace{80pt} \times  f_{eq}(\epsilon) L_0^{\beta\gamma}(\epsilon'') \frac{\partial}{\partial\epsilon} f_{eq}(\epsilon-\epsilon'') \nn
&&\hspace{30pt} + K_0^{\alpha\beta}(\epsilon') f_{eq}(\epsilon) f(\epsilon-\epsilon') L_0^{\beta\gamma}(\epsilon'') \nn
&&\hspace{80pt} \times\frac{\partial}{\partial \epsilon} f_{eq}(\epsilon-\epsilon'-\epsilon'') \bigg] E_\gamma(\omega).
\label{Eq:Gamma}
\end{eqnarray}
It turns out that the last two terms proportional to $K_1 L_0$ and $K_0 L_0$ are higher order in disorder strength $\lambda$. In the present study we will not take into account these contributions.

Inserting Eq. (\ref{Eq:Gamma}) into the formal expression of the electrical current, we obtain the electrical conductivity in the absence of external magnetic fields
\begin{eqnarray}
\frac{\sigma(\omega)}{\sigma_0}
&=& \frac{1}{1 - i\omega\tau + \alpha(\omega)\tau} \nn
&&+ \frac{2}{3} \frac{1}{\big[ 1 - i\omega\tau + \alpha(\omega)\tau \big]^2} \int\!\frac{d\epsilon'}{\pi} \ \frac{\partial}{\partial \epsilon'} \left( \epsilon' \coth \frac{\epsilon'}{2T} \right) \nn
&&\times \bigg\{ K_0(\epsilon') - K_1(\epsilon') - \frac{3}{2} \big[ 1-i\omega\tau + \alpha(\omega)\tau \big] \frac{L_0(\epsilon')}{v_F\tau} \bigg\}. \nn
\label{eq:Cond}
\end{eqnarray}
Here, we have
\beqns
K_0(\epsilon')
&\approx& -\frac{\tau^2}{4\pi^2 N_F} \IMM \int_0^\infty\!dq q^2 \frac{1}{1-i\epsilon'\tau + {3\over2}D_0q^2\tau} \nn
&&\hspace{40pt}\times \left( \frac{1}{2} + \frac{1}{-i\epsilon'\tau + {3\over2}D_0q^2\tau} \right), \\
K_1(\epsilon')
&\approx& \frac{\tau^2}{4\pi^2 N_F} \IMM \int_0^\infty\!dq q^2 \frac{1}{-i\epsilon'\tau + {3\over2}D_0q^2\tau} \nn
&&\hspace{30pt} \times\left( \frac{1}{2} + \frac{1}{1-i\epsilon'\tau + {3\over2}D_0q^2\tau} \right), \\
\frac{L_0(\epsilon')}{v_F\tau}
&\approx& \frac{\tau^2}{2\pi^2 N_F} \IMM \int_0^\infty\!dq q^2 \frac{1}{-i\epsilon'\tau + {3\over2}D_0q^2\tau} \nn
&&\hspace{-30pt} \times\frac{1}{\big( 1-i\epsilon'\tau + {3\over2}D_0q^2\tau \big)^3} \left( \frac{3}{2} + \frac{1}{-i\epsilon'\tau + {3\over2}D_0q^2\tau} \right).\nn
\eeqns

Introducing dimensionless quantities of $\tilde q = ql$, $\tilde \epsilon' = \epsilon/E_F$, $\tilde T = T/E_F$, and $\lambda = 1/(\pi k_Fl)$, we simplify the dc conductivity ($\omega = 0$) as
\begin{eqnarray}
&&\frac{\sigma(0)}{\sigma_0}
= \frac{1}{1+\alpha(0)\tau} \nn
&&- \frac{1}{1+\alpha(0)\tau} \frac{\lambda}{2} \int_0^{\infty}\!d\tilde\epsilon' \frac{\partial}{\partial\tilde\epsilon'} \left( \tilde\epsilon' \tanh \frac{\tilde\epsilon'}{2\tilde T} \right) \nn
&& \times \IMM \int_0^\infty\!d\tilde q \tilde q^2 \frac{1}{\big( -i{\tilde\epsilon' \over 2\pi\lambda} + {1\over2} \tilde q^2 \big) \left( 1-i{\tilde\epsilon' \over 2\pi\lambda} + {1\over2} \tilde q^2 \right)^3} \nn
&&\hspace{100pt} \times \left( \frac{3}{2} + \frac{1}{-i{\tilde\epsilon' \over 2\pi\lambda} + {1\over2} \tilde q^2} \right) \nn
&&- \frac{1}{\big( 1+\alpha(0)\tau \big)^2} \frac{\lambda}{6} \int_0^1\!d\tilde\epsilon' \frac{\partial}{\partial\tilde\epsilon'} \left( \tilde\epsilon' \tanh \frac{\tilde\epsilon'}{2\tilde T} \right) \nn
&&\times \IMM \int_0^\infty \!d\tilde q \tilde q^2 \frac{1}{1-i{\tilde\epsilon' \over 2\pi\lambda} + {1\over2} \tilde q^2} \left( \frac{1}{2} + \frac{1}{-i{\tilde\epsilon' \over 2\pi\lambda} + {1\over2} \tilde q^2} \right) \nn
&&- \frac{1}{\big( 1+\alpha(0)\tau \big)^2} \frac{\lambda}{6} \int_0^1\!d\tilde\epsilon' \frac{\partial}{\partial\tilde\epsilon'} \left( \tilde\epsilon' \tanh \frac{\tilde\epsilon'}{2\tilde T} \right) \nn
&&\times \IMM \int_0^\infty\!d\tilde q \tilde q^2 \frac{1}{-i{\tilde\epsilon' \over 2\pi\lambda} + {1\over2} \tilde q^2} \left( \frac{1}{2} + \frac{1}{1-i{\tilde\epsilon' \over 2\pi\lambda} + {1\over2} \tilde q^2} \right), \nn
\end{eqnarray}
where $\alpha(0)\tau = \frac{1}{\pi N_F} \int'\!\frac{d^3q}{(2\pi)^3} \frac{1}{D_0q^2} = 3\pi\lambda^2$. We note that the integration limit of the last two integrals is given by $0<\tilde\epsilon'<1$ (or $0<\epsilon'<E_F$). In the zero-temperature limit, we see ${\partial \over \partial\tilde\epsilon'} \left( \tilde\epsilon' \coth {\tilde\epsilon' \over 2\tilde T} \right) \rightarrow 1$. Then, the integrals can be performed exactly, which yields
\begin{eqnarray}
\frac{\sigma(0)}{\sigma_0}
&=& \frac{1}{1+3\pi\lambda^2} - \frac{3\sqrt{2}\pi^2}{8} \frac{\lambda^2}{1+3\pi\lambda^2} \nn
&&- \frac{\sqrt{2}\pi^2}{9} \frac{\lambda^2}{\big( 1 + 3\pi\lambda^2 \big)^2} \mathcal{K}_0 \left( {1\over 2\pi\lambda} \right) \nn
&&- \frac{2\sqrt{2}\pi^2}{9} \frac{\lambda^2}{\big( 1 + 3\pi\lambda^2 \big)^2}  \mathcal{K}_1\left( {1\over 2\pi\lambda} \right),
\label{eq:condZT}
\end{eqnarray}
where
\begin{subequations}
\begin{eqnarray}
\mathcal{K}_0(y) &=& \big( 1+y^2 \big)^{3\over4} \cos\left( {3\over2} \tan^{-1}y \right) + \sqrt{2} y^{3\over2} - 1, \\
\mathcal{K}_1(y) &=& \big( 1+y^2 \big)^{3\over4} \cos\left( {3\over2} \tan^{-1}y \right) + {3\sqrt{2}\over4} y^{3\over2} - 1. \nn
\end{eqnarray}
\label{eq:K}
\end{subequations}
Note that $\mathcal{K}_0(0) = \mathcal{K}_1(0) = 0$. For $T/E_F \ll 1$ or $T\tau \ll 1/(2\pi\lambda)$, we perform the integrals rather reliably even at finite temperatures and obtain
\begin{eqnarray}
\frac{\sigma(0)}{\sigma_0}
&=& \frac{1}{1+3\pi\lambda^2}
- \frac{3\sqrt{2}\pi^2}{8} \frac{\lambda^2}{1+3\pi\lambda^2} \mathcal{L}_0 \big( 3T\tau \big) \nn
&& - \frac{\sqrt{2}\pi^2}{9} \frac{\lambda^2}{\big( 1 + 3\pi\lambda^2 \big)^2} \left\{ \mathcal{K}_0 \left( {1\over 2\pi\lambda} \right) - \mathcal{K}_0 \big( 3T\tau \big) \right\} \nn
&&- \frac{2\sqrt{2}\pi^2}{9} \frac{\lambda^2}{\big( 1 + 3\pi\lambda^2 \big)^2}  \left\{ \mathcal{K}_1 \left( {1\over 2\pi\lambda} \right) - \mathcal{K}_1 \big( 3T\tau \big) \right\}, \nn
\end{eqnarray}
where
\begin{eqnarray}
\mathcal{L}_0(y)
&=& 2(1+y^2)^{3\over4} \cos \left( {3\over2} \tan^{-1} y \right) + \frac{4\sqrt{2}}{3}  y^{3\over2} \nn
&& - \frac{1 + {2\over3} y^2}{\big( 1+y^2 \big)^{1\over4}} \cos\left( {1\over2} \tan^{-1} y \right) - \frac{4\sqrt{2}}{3} \sqrt{y} .
\label{eq:L}
\end{eqnarray}
Note that $\mathcal{L}_0(0)= 1$ and $\mathcal{L}_0(\infty) = 0$.

In order to investigate the many-body localization transition, we should take into account the region of strong disorder strengths, i.e., $\lambda \gg 1$. Unfortunately, it is difficult to justify the above expression in the region of strong disorder strengths. It is easy to see that interaction contributions become enhanced as $\lambda$ increases in Eq. (\ref{eq:condZT}), which causes the conductivity to be negative. In this respect we need to resum such interaction effects in the RPA fashion. We rewrite Eq. (\ref{eq:Cond}) as follows
\begin{eqnarray}
&&\frac{\sigma(0)}{\sigma_0}
= \Bigg[ 1 + \alpha(0)\tau - \frac{2}{3} \int\!\frac{d\epsilon'}{\pi} \ \frac{\partial}{\partial \epsilon'} \left( \epsilon' \coth \frac{\epsilon'}{2T} \right) \nn
&&\times \bigg\{ K_0(\epsilon') - K_1(\epsilon') - \frac{3}{2} \big[ 1 + \alpha(0)\tau \big] \frac{L_0(\epsilon')}{v_F\tau} \bigg\} \Bigg]^{-1}.
\end{eqnarray}
%
%

\begin{figure}[t!]
\includegraphics[width=8.5cm]{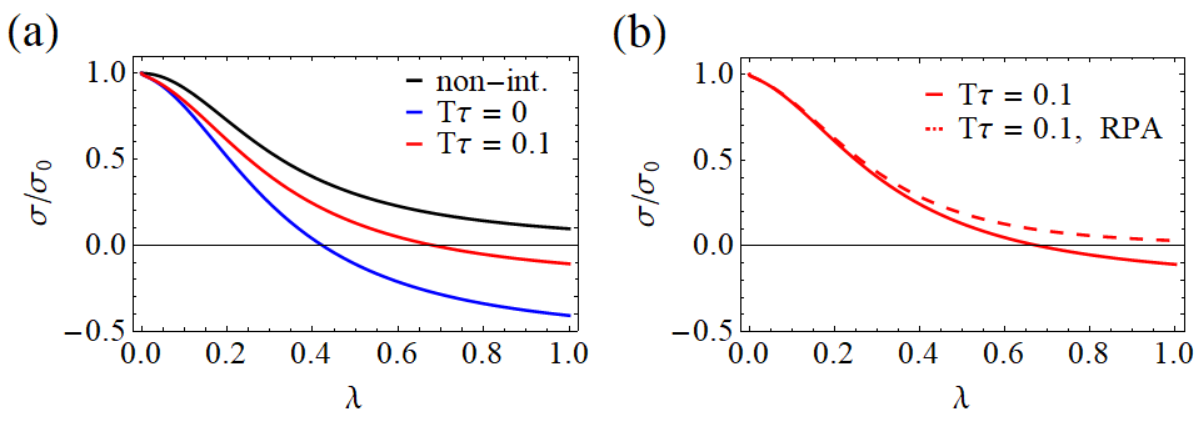}
\caption{Electrical conductivity in terms of disorder strength $\lambda$. (a) The black line represents electric conductivity in the absence of electron correlations. The blue and red lines express the electric conductivity with interactions for $T\tau = 0$ and $T\tau = 0.1$, respectively. (b) We compare the non-RPA conductivity (red-solid) with an RPA (red-dashed) form for interactions at $T\tau = 0.1$.}
\label{fig04}
\end{figure}

We show the electrical conductivity as a function of the disorder strength in Fig. \ref{fig04}. Figure \ref{fig04} (a) displays that the electrical conductivity becomes negative when the disorder strength exceeds a certain value. This implies that the weak-disorder expansion does not work in the region of strong disorder strengths. Figure \ref{fig04} (b) shows that the problem of the negative conductivity is cured in the RPA resummation.

\section{A self-consistent theory for a many-body localization insulator-metal transition}

In order to prepare for the self-consistent equation of the diffusion coefficient, we take into account the following replacement
\begin{eqnarray}
v_F^2q^2 \rightarrow \frac{3Dq^2}{\tau} ,
\end{eqnarray}
based on the Einstein relation, where $D=D(\omega=0)$ is the renormalized diffusion coefficient. This construction allows renormalization of the diffusive dynamics in a self-consistent way. Then, the interaction kernels read
\beqns
K_0(\epsilon')
&\approx& -\frac{\tau^2}{4\pi^2 N_F} \IMM \int_0^\infty\!dq q^2 \frac{1}{1-i\epsilon'\tau + {3\over2}Dq^2\tau} \nn
&&\hspace{40pt} \times\left( \frac{1}{2} + \frac{1}{-i\epsilon'\tau + {3\over2}Dq^2\tau} \right), \\
K_1(\epsilon')
&\approx& \frac{\tau^2}{4\pi^2 N_F} \IMM \int_0^\infty\!dq q^2 \frac{1}{-i\epsilon'\tau + {3\over2}Dq^2\tau} \nn
&&\hspace{30pt} \times\left( \frac{1}{2} + \frac{1}{1-i\epsilon'\tau + {3\over2}Dq^2\tau} \right), \\
\frac{L_0(\epsilon')}{v_F\tau}
&\approx& \frac{\tau^2}{2\pi^2 N_F} \IMM \int_0^\infty\!dq q^2 \frac{1}{-i\epsilon'\tau + {3\over2}Dq^2\tau} \nn
&&\times\frac{1}{\big( 1-i\epsilon'\tau + {3\over2}Dq^2\tau \big)^3} \left( \frac{3}{2} + \frac{1}{-i\epsilon'\tau + {3\over2}Dq^2\tau} \right). \nn
\eeqns

\begin{figure*}[t!]
\includegraphics[width=18cm]{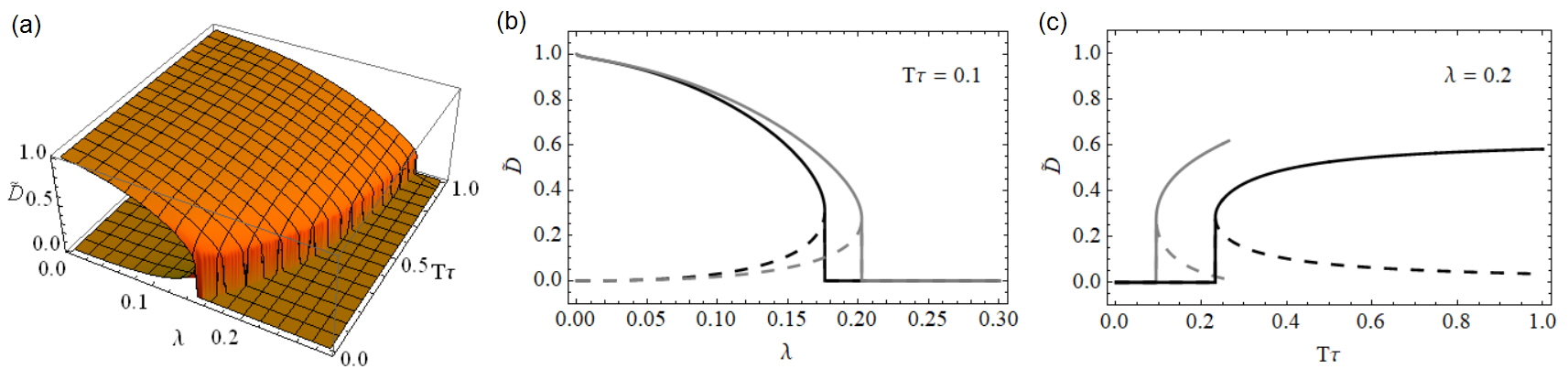}
\caption{Diffusion coefficient as a function of the disorder strength $\lambda = 1 / (\pi k_{F} l)$ and reduced temperature $T\tau$ in our Boltzmann-transport self-consistent theory for a many-body localization transition. (a) expresses a three-dimensional plot of the diffusion coefficient as a function of both disorder strength and temperature. Note that there are two surfaces in a metallic phase. (b) displays the diffusion coefficient as a function of the disorder strength at a finite temperature $\tilde T = 0.1$. The grey-solid (grey-dashed) line marks the larger (smaller) solution of Eq. (\ref{MBL_Analytic_Expression}). A deviation between the numerical analysis and the analytic approach originates from the fact that we neglect the low-temperature contribution given by $\int_{0}^{\tilde{T}} d \tilde{\epsilon}$ in the analytic approach. (c) shows the diffusion coefficient as a function of temperature in a many-body localized insulating phase of $\lambda = 0.2$. The reason why we cut the grey-solid (grey-dashed) line is that the analytic expression Eq. (\ref{MBL_Analytic_Expression}) can be justified within the regime of $T\tau \ll 1/(2\pi\lambda)$.}
\label{fig05}
\end{figure*}

Considering dimensionless parameters of $\tilde D = D/D_0$ and $\tilde q = ql$, we rewrite the self-consistent equation in an RPA fashion
\begin{eqnarray}
&&\tilde D
= \Bigg[ 1+\alpha_{SC}(0)\tau + \big( 1+\alpha_{SC}(0)\tau \big) \frac{\lambda}{2}  \nn
&&\times\int_0^{\infty}\!d\tilde\epsilon' \frac{\partial}{\partial\tilde\epsilon'} \left( \tilde\epsilon' \tanh \frac{\tilde\epsilon'}{2\tilde T} \right) \IMM \int_0^\infty\!d\tilde q \tilde q^2 \frac{1}{-i{\tilde\epsilon' \over 2\pi\lambda} + {1\over2} \tilde D \tilde q^2} \nn
&&\hspace{30pt}\times \frac{1}{\left( 1-i{\tilde\epsilon' \over 2\pi\lambda} + {1\over2} \tilde D \tilde q^2 \right)^3} \left( \frac{3}{2} + \frac{1}{-i{\tilde\epsilon' \over 2\pi\lambda} + {1\over2} \tilde D \tilde q^2} \right) \nn
&&+ \frac{\lambda}{6} \int_0^1\!d\tilde\epsilon' \frac{\partial}{\partial\tilde\epsilon'} \left( \tilde\epsilon' \tanh \frac{\tilde\epsilon'}{2\tilde T} \right) \IMM \int_0^\infty \!d\tilde q \tilde q^2 \frac{1}{1-i{\tilde\epsilon' \over 2\pi\lambda} + {1\over2} \tilde D \tilde q^2} \nn
&&\hspace{80pt} \times \left( \frac{1}{2} + \frac{1}{-i{\tilde\epsilon' \over 2\pi\lambda} + {1\over2} \tilde D \tilde q^2} \right) \nn
&&+ \frac{\lambda}{6} \int_0^1\!d\tilde\epsilon' \frac{\partial}{\partial\tilde\epsilon'} \left( \tilde\epsilon' \tanh \frac{\tilde\epsilon'}{2\tilde T} \right) \IMM \int_0^\infty\!d\tilde q \tilde q^2 \frac{1}{-i{\tilde\epsilon' \over 2\pi\lambda} + {1\over2} \tilde D \tilde q^2} \nn
&&\hspace{80pt} \times \left( \frac{1}{2} + \frac{1}{1-i{\tilde\epsilon' \over 2\pi\lambda} + {1\over2} \tilde D \tilde q^2} \right) \Bigg]^{-1},
\label{Self_Consisteny_Resummation_MBL}
\end{eqnarray}
where $\alpha_{SC}(0)\tau = \frac{1}{\pi N_F} \int'\!\frac{d^3q}{(2\pi)^3} \frac{1}{Dq^2} = 3\pi\lambda^2/\tilde D$. The subscript $SC$ means ``self-consistency".

Following the previous subsection, we perform the integrals in the interaction kernels for $T=0$ and $T\tau \ll 1/(2\pi\lambda)$. As a result, we find
\begin{eqnarray}
&&\tilde D
= \Bigg[ \left( 1+ \frac{3\pi\lambda^2}{\tilde D} \right) \left( 1 + \frac{3\sqrt{2}\pi^2}{8} \frac{\lambda^2}{\tilde D^{3\over2}} \right) \nn
&&+ \frac{\sqrt{2}\pi^2}{9} \frac{\lambda^2}{\tilde D^{3\over2}} \mathcal{K}_0 \left( {1\over 2\pi\lambda} \right)
+ \frac{2\sqrt{2}\pi^2}{9} \frac{\lambda^2}{\tilde D^{3\over2}} \mathcal{K}_1 \left( {1\over 2\pi\lambda} \right) \Bigg]^{-1}
\end{eqnarray}
for $T=0$ and
\begin{eqnarray}
\tilde D
&=& \Bigg[ \left( 1+\frac{3\pi\lambda^2}{\tilde D} \right) \left( 1 + \frac{3\sqrt{2}\pi^2}{8} \frac{\lambda^2}{\tilde D^{3\over2}} \mathcal{L}_0 \big(3T\tau \big) \right) \nn
&&+ \frac{\sqrt{2}\pi^2}{9} \frac{\lambda^2}{\tilde D^{3\over2}} \left\{ \mathcal{K}_0 \left( {1\over 2\pi\lambda} \right) - \mathcal{K}_0 \big( 3T\tau \big) \right\} \nn
&&+ \frac{2\sqrt{2}\pi^2}{9} \frac{\lambda^2}{\tilde D^{3\over2}}  \left\{ \mathcal{K}_1 \left( {1\over 2\pi\lambda} \right) - \mathcal{K}_1 \big( 3T\tau \big) \right\} \Bigg]^{-1} \label{MBL_Analytic_Expression}
\end{eqnarray}
for $T\tau \ll 1/(2\pi\lambda)$, respectively. The functions $\mathcal{L}_0(y)$, $\mathcal{K}_0(y)$ and $\mathcal{K}_1(y)$ are given in Eq. (\ref{eq:K}) and Eq. (\ref{eq:L}). Taking the limit of $\tilde{D} \rightarrow 1$ in the right hand side of these equations, we recover those of the previous section in the RPA form.

\begin{figure}[t!]
\includegraphics[width=7cm]{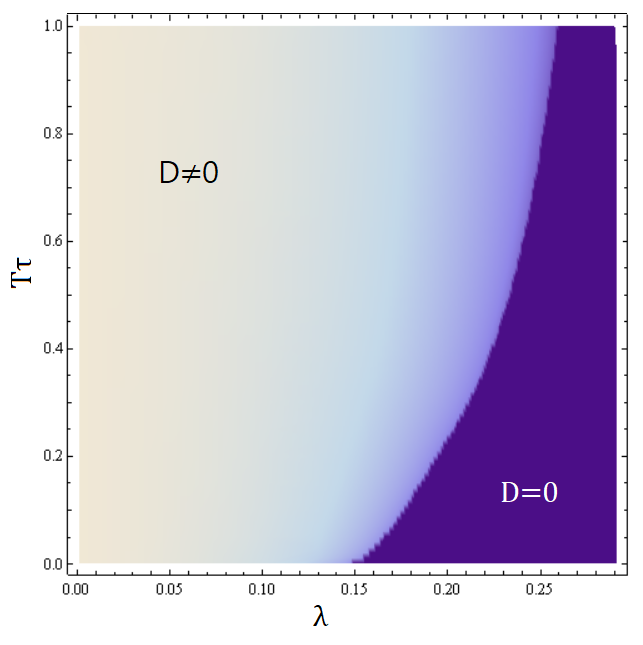}
\caption{Phase diagram of a many-body localization transition in the plane of the disorder strength and temperature.}
\label{fig06}
\end{figure}

We solve Eq. (\ref{Self_Consisteny_Resummation_MBL}) numerically and find the diffusion coefficient as a function of the disorder strength $\lambda$ and temperature $T$. The three-dimensional plot of Fig. \ref{fig05} (a) shows the diffusion coefficient as a function of the disorder strength $\lambda$ and temperature $T\tau$. We find that there are two solutions in a metallic phase, which results from the introduction of self-consistency in the RPA expression. The larger solution reproduces $\tilde D = 1$ in the clean limit ($\lambda=0$) while the smaller one vanishes. We emphasize again that only one solution exists in the clean limit. Mathematically, the smaller solution results from the $\tilde D\tilde q^2$ combination and the spatial dimension $d=3$ in interaction kernels. These lead to the $\tilde D^{-{3\over2}}$ dependence of interaction kernels and dominate the weak localization term in small $\tilde D$, giving rise to the small solution in Eq. (\ref{Self_Consisteny_Resummation_MBL}). We point out that dominant contributions occur from $L_{0}$ among interaction kernels in the vicinity of the metal-insulator transition, responsible for dephasing effects \cite{Interaction_Boltzmann_Transport_Theory}. As the disorder strength increases from the metallic side, these two solutions get closer and merge together at $\lambda = \lambda_c$ with a finite $\tilde D = \tilde D_c$. Increasing $\lambda$ further, $\tilde D$ drops to zero in a discontinuous fashion, which suggests the first order metal-insulator transition. See Fig. \ref{fig05} (b). When temperature increases, $\lambda_c$ is enhanced and $\tilde D_c$ is reduced. We point out that interaction effects vanish in the $T\tau \rightarrow \infty$ limit. In this case the transition belongs to the same universality class as that of the Wolfle-Vollhardt theory, given by $\lambda_c = 1/{\sqrt{3\pi}}$ and $\tilde D = 0$. The discontinuous change of the electrical conductivity turns into the continuous evolution of the Wolfle-Vollhardt theory, which leads us to suspect that the nature of this insulator-metal transition changes from the first order to the second one in the $T\tau \rightarrow \infty$ limit, identified with a multicritical point. Fixing the disorder strength $\lambda$ around the $\lambda_c$, we also find an insulator-to-metal transition of the first order in temperature, as shown in Fig. \ref{fig05}(c). A phase diagram is drawn in Fig. \ref{fig06} in the plane of the disorder strength and temperature, which describes a quantum phase transition from a many-body localized state to a metallic phase.

The existence of the smaller solution for the diffusion coefficient in a metallic phase suspects the reliability of the present theoretical framework. It can be an artifact of our RPA-based self-consistent formulation. In other words, higher-order interaction corrections should be taken into account beyond the present description for interaction kernels, where possible interplays between disorders and interactions may renormalize electron correlations stronger. In particular, the system of our interests can become inhomogeneous due to such renormalization effects. We recall that our distribution function is based on the assumption of homogeneity of the system. Actually, the inhomogeneity occurs in one dimension quite often. On the other hand, it is not easy for the system to be extremely inhomogeneous in three dimensions. Ref. \cite{Anomalous_Diffusion} has shown that anomalous diffusion should be taken into account in order to make the Wolfle-Vollhardt self-consistent equation be consistent with the scaling theory near the Anderson metal-insulator transition \cite{Altshuler_Aronov}. Following Ref. \cite{Anomalous_Diffusion}, we considered anomalous diffusions given by $D q^{\eta_{D}}$, where $\eta_{D}$ is an anomalous scaling dimension. Unfortunately, we could not find a physically appealing solution with a nontrivial $\eta_{D}$, i.e., $\eta_{D} \not= 0$.

Resorting to the coexistence of two types of diffusion coefficients, it is natural to conclude that the distribution function of the diffusion coefficient would be given by a bimodal function, which has a two peak-like feature as a function of the diffusion coefficient. We believe that the emergence of this bimodal distribution function at least in the vicinity of the many-body localization transition can be either verified or falsified in the numerical simulation. Recently, we performed a Hartree-Fock study for interacting electrons with randomness, where Coulomb interactions are treated within the Hartree-Fock approximation, but disorder effects are taken into account exactly \cite{HF_Study_Lee_Kim}. Here, we focused on a less disordered regime below a critical value of disorder. We may repeat the same study but above the critical disorder strength where all electrons are localized. Calculating the electrical conductivity for various realizations of disorders, we can find the distribution function for the diffusion coefficient as a function of temperature.

\section{Summary}

In summary, generalizing the Hershfield-Ambegaokar Boltzmann transport theory \cite{Boltzmann_Transport_Theory_AMIT} based on the study of Zala-Narozhny-Aleiner \cite{Interaction_Boltzmann_Transport_Theory}, we extended the Wolfle-Vollhardt self-consistent equation \cite{Self_Consistent_Theory_AMIT} for the diffusion coefficient in the presence of electron correlations, where not only Altshuler-Aronov corrections \cite{Altshuler_Aronov} but also dephasing effects \cite{Dephasing} are taken into account. As a result, we find that a many-body localized insulating state at low temperatures turns into a metallic phase at high temperatures due to dephasing effects. This insulator-metal transition shows the first order in three dimensions, rather unexpected. The origin of this first order phase transition is not clear. Inhomogeneity of the diffusion coefficient may have to be introduced into the generalized Boltzmann transport theory beyond the present level of approximation.

\begin{acknowledgments}
This study was supported by the Ministry of Education, Science, and Technology (No. NRF-2015R1C1A1A01051629 and No. 2011-0030046) of the National Research Foundation of Korea (NRF). We would like to thank M. S. Foster for his correspondence.
\end{acknowledgments}

\appendix

\begin{widetext}
\section{Angle averages in three dimensions}

The equation for $D$ is given by
\begin{eqnarray}
\left( -i\omega + \frac{1}{\tau} + iv_F \bm n \cdot \bm q \right) D(\bm n, \bm n'; \omega, \bm q) - \frac{1}{\tau} \big< D(\bm n, \bm n'; \omega, \bm q) \big>_n = 4\pi \delta(\bm n, \bm n'),
\end{eqnarray}
or
\begin{eqnarray}
D(\bm n, \bm n') - \frac{1}{\tau} D_0(\bm n) \big< D(\bm n, \bm n') \big>_n = 4\pi D_0(\bm n) \delta(\bm n, \bm n'), \ \ \ \ \ \ D_0(\bm n) \equiv \frac{1}{-i\omega + \frac{1}{\tau} + iv_F \bm n \cdot \bm q}.
\end{eqnarray}
Averaging both sides over $\bm n$, we obtain
\begin{eqnarray}
\big< D(\bm n, \bm n') \big>_n - \frac{1}{\tau} \big< D_0(\bm n) \big>_n \big< D(\bm n, \bm n') \big>_n = D_0(\bm n') \ \ \ \Rightarrow \ \ \
\big< D(\bm n, \bm n') \big>_n = \frac{1}{1-\frac{1}{\tau} \big< D_0(\bm n) \big>_n} D_0(\bm n').
\end{eqnarray}
As a result, we find
\begin{eqnarray}
D(\bm n, \bm n') = 4\pi D_0(\bm n) \delta(\bm n, \bm n') + \frac{1}{\tau - \big< D_0(\bm n) \big>_n} D_0(\bm n) D_0(\bm n').
\end{eqnarray}

Recalling the definition of the angle average in three dimensions
\begin{eqnarray}
\Big< \cdots \Big>_n \equiv \int\!\frac{d\Omega}{4\pi} \ \Big( \cdots \Big) = \frac{1}{2}\int_{-1}^1\!d(\cos\theta) \frac{1}{2\pi}\int_0^{2\pi}\!d\phi \ \Big( \cdots \Big), \ \ \ \ \ \
\bm n = (\sin\theta\cos\phi, \sin\theta\sin\phi, \cos\theta) ,
\end{eqnarray}
we obtain angle averages of all quantities that are necessary for the present study
\beqns
\Big< D_0(\bm n) \Big>_n
&=& \frac{1}{4\pi} \int_{-1}^1\!d\cos\theta \int_0^{2\pi}\!d\phi \ \frac{1}{A + iv_Fq\cos\theta}
= \frac{1}{2iv_F q} \log \frac{A + iv_Fq}{A - iv_Fq}
\equiv \frac{1}{C}, \\
\Big< n_\alpha D_0(\bm n) \Big>_n
&=& \frac{1}{4\pi} \int_{-1}^1\!d\cos\theta \int_0^{2\pi}\!d\phi \ \frac{n_\alpha}{A + iv_Fq\cos\theta}
= \frac{q_\alpha}{iv_Fq^2} \bigg( 1 - \frac{A}{C} \bigg), \\
\Big< n_\alpha n_\beta D_0(\bm n) \Big>_n
&=& \delta_{\alpha\beta} \bigg[ \cos^2\theta_q \Big< n^2 D_0(\bm n) \Big>_{q_\parallel} + \sin^2\theta_q \Big< n^2 D_0(\bm n) \Big>_{q_\perp} \bigg], \\
\Big< n^2 D_0(\bm n) \Big>_{q_\parallel}
&=& \frac{A}{v_F^2q^2} \bigg( 1 - \frac{A}{C} \bigg), \ \ \
\Big< n^2 D_0(\bm n) \Big>_{q_\perp}
= -\frac{A}{v_F^2q^2} \bigg( 1 - \frac{A^2+ v_F^2q^2}{AC} \bigg), \nn
\Big< D(\bm n, \bm n') \Big>_{n}
&=& \frac{C\tau}{C\tau - 1} D_0(\bm n'), \\
\Big< n_\alpha D(\bm n, \bm n') \Big>_n
&=& n'_\alpha D_0(\bm n') + \Big< n_\alpha D_0(\bm n) \Big>_n D_0(\bm n') \frac{C}{C\tau -1} \nn
&=& n'_\alpha D_0(\bm n') + \frac{q_\alpha}{iv_Fq^2} \bigg( 1 - \frac{A}{C} \bigg) \frac{C}{C\tau -1} D_0(\bm n'), \\
\Big< D(\bm n, \bm n') \Big>_{n,n'}
&=& \frac{C\tau}{C\tau - 1} \Big< D_0(\bm n') \Big>_{n'}
= \frac{\tau}{C\tau - 1}, \\
\Big< n_\alpha D(\bm n, \bm n') \Big>_{n,n'}
&=& \Big< n'_\alpha D_0(\bm n') \Big>_{n'} + \frac{q_\alpha}{iv_Fq^2} \bigg( 1 - \frac{A}{C} \bigg) \frac{C}{C\tau -1} \Big< D_0(\bm n') \Big>_{n'}
= \frac{q_\alpha}{iv_Fq^2} \bigg( 1 - \frac{A}{C} \bigg) \frac{C\tau}{C\tau -1}, \\
\Big< D(\bm n, \bm n') n'_\alpha  \Big>_{n,n'}
&=& \frac{q_\alpha}{iv_Fq^2} \bigg( 1 - \frac{A}{C} \bigg) \frac{C\tau}{C\tau -1}, \\
\Big< n_\alpha D(\bm n, \bm n') n'_\beta  \Big>_{n,n'}
&=& \Big< \Big< n_\alpha D(\bm n, \bm n') \Big>_n n'_\beta  \Big>_{n'}
= \Big< n'_\alpha n'_\beta D_0(\bm n') \Big>_{n'} + \frac{q_\alpha}{iv_Fq^2} \bigg( 1 - \frac{A}{C} \bigg) \frac{C}{C\tau -1} \Big< n'_\beta D_0(\bm n') \Big>_{n'} \nn
&=& \delta_{\alpha\beta} \bigg[ \cos^2\theta_q \Big< n^2 D_0(\bm n) \Big>_{q_\parallel} + \sin^2\theta_q \Big< n^2 D_0(\bm n) \Big>_{q_\perp} \bigg] - \frac{q_\alpha q_\beta}{v_F^2q^4} \bigg( 1 - \frac{A}{C} \bigg)^2 \frac{C}{C\tau -1} ,
\eeqns
where $A\equiv i\omega + 1/\tau$.

\section{Angle averages in two dimensions}

We summarize angle averages for various quantities in two dimensions as follows
\beqns
\Big< D_0(\bm n) \Big>_n
&=& \int_0^{2\pi}\!\frac{d\theta}{2\pi} \ \frac{1}{A + iv_Fq \cos\theta}
= \frac{1}{\sqrt{A^2 + v_F^2 q^2}} \equiv \frac{1}{C}, \\
\Big< n_\alpha D_0(\bm n) \Big>_n
&=& \int_0^{2\pi}\!\frac{d\theta}{2\pi} \ \frac{(\cos\theta, \sin\theta)}{A + iv_Fq \cos\theta}
= \frac{1}{iv_Fq} \frac{C-A}{C}, 0 \nn
&\rightarrow& \frac{q_\alpha}{iv_Fq^2} \frac{C-A}{C}, \\
\Big< n_\alpha n_\beta D_0(\bm n) \Big>_n
&=& \delta_{\alpha\beta} \int\!\frac{d\theta}{2\pi} \ \frac{(\cos^2\theta, \sin^2\theta)}{A + iv_F q \cos\theta}
= \delta_{\alpha\beta} \frac{1}{v_F^2 q^2} \frac{A(C-A)}{C}, \delta_{\alpha\beta} \frac{1}{v_F^2q^2} \frac{v_F^2q^2 - A(C-A)}{C} \nn
&\rightarrow& \begin{pmatrix} D_{p_\parallel} \cos^2 \alpha + D_{p_\perp} \sin^2\alpha & (D_{p_\parallel} - D_{p_\perp}) \cos\alpha\sin\alpha \\ (D_{p_\parallel} - D_{p_\perp}) \cos\alpha\sin\alpha &  D_{p_\parallel} \cos^2 \alpha + D_{p_\perp} \sin^2\alpha  \end{pmatrix}, \nn
&&D_{p_\parallel} = \frac{1}{v_F^2q^2} \frac{A(C-A)}{C}, \ \ \
D_{p_\perp} = \frac{1}{v_F^2 q^2} \frac{v_F^2q^2 - A(C-A)}{C}, \\
\Big< D_0(\bm n) n_\alpha D_0(\bm n) \Big>_n
&=& -\frac{iv_F q_\alpha}{C^3}, \nn
\Big< D_0(\bm n) n_\alpha \frac{\partial}{\partial q_\beta} D_0(\bm n) \Big>_n
&=& - \frac{iv_F q_\alpha q_\beta}{2q^2} \frac{A^2 - 2v_F^2 q^2}{C^5}, \\
\Big< D(\bm n, \bm n') n'_\alpha \frac{\partial}{\partial q_\beta} D(\bm n', \bm n'') \Big>_{n,n',n''}
&=& \frac{1}{2} \frac{\partial}{\partial q_\beta} \Big< D(\bm n, \bm n') n'_\alpha D(\bm n', \bm n'') \Big>_{n,n',n''} \nn
&=& \frac{1}{2} \frac{\partial}{\partial q_\beta} \bigg[ \left( \frac{C\tau}{C\tau -1} \right)^2 \left( -iv_F \frac{q_\alpha}{C^3} \right) \bigg] \nn
&=& -\frac{iv_F}{2} \Bigg[ \delta_{\alpha\beta} \frac{\big( C\tau \big)^2}{C^3 \big( C\tau -1 \big)^2} - v_F^2 q_\alpha q_\beta \frac{\tau^2 \big( 3C\tau - 1 \big)}{C^3 \big( C\tau -1\big)^3} \Bigg], \\
\Big< D(\bm n, \bm n') \Big>_n
&=& \frac{C\tau}{C\tau -1} D_0(\bm n'), \\
\Big< n_\alpha D(\bm n, \bm n') \Big>_n
&=& n_\alpha' D_0(\bm n') + \Big< n_\alpha D_0(\bm n) \Big>_n D_0(\bm n') \frac{C}{C\tau -1} \nn
&=& n_\alpha' D_0(\bm n') + D_0(\bm n') \frac{q_\alpha}{iv_F q^2} \frac{C-A}{C\tau -1}, \\
\Big< D(\bm n, \bm n') \Big>_{n,n'}
&=& \frac{C\tau}{C\tau -1} \Big< D_0(\bm n') \Big>_{n'}
= \frac{\tau}{C\tau -1}, \\
\Big< n_\alpha D(\bm n, \bm n') \Big>_{n,n'}
&=& \Big< n_\alpha' D_0(\bm n') \Big>_{n'} + \Big< D_0(\bm n') \Big>_{n'} \frac{q_\alpha}{iv_F q^2} \frac{C-A}{C\tau -1} \nn
&=& \frac{q_\alpha}{iv_Fq^2} \frac{C-A}{C} + \frac{q_\alpha}{iv_F q^2} \frac{C-A}{C(C\tau -1)}
= \frac{q_\alpha}{iv_Fq^2} \frac{(C-A)\tau}{C\tau -1}, \\
\Big< n_\alpha D(\bm n, \bm n') n'_\beta  \Big>_{n,n'}
&=& \Big< \Big< n_\alpha D(\bm n, \bm n') \Big>_n n'_\beta  \Big>_{n'}
= \Big< n'_\alpha n'_\beta D_0(\bm n') \Big>_{n'} + \frac{q_\alpha}{iv_Fq^2} \frac{C-A}{C\tau -1} \Big< n'_\beta D_0(\bm n') \Big>_{n'} \nn
&=&\Big< n'_\alpha n'_\beta D_0(\bm n') \Big>_{n'} - \frac{q_\alpha q_\beta}{v_F^2q^4} \frac{(C-A)^2}{C(C\tau -1)} ,
\eeqns
based on the coordinate representation of $\bm q = q \hat{\bm x}$ and $\bm n = (\cos\theta, \sin\theta)$.
\end{widetext}


\begin{thebibliography}{9}
\bibitem{Anomalous_Diffusion} This correlation-length critical exponent is not consistent with the conventional wisdom of $\nu = \frac{1}{2} + \frac{1}{d - 2}$, where we have $\nu = \frac{1}{2} + 1$ in $d = 3$. It has been proposed that this problem may be resolved if one introduces anomalous diffusion into the Wolfle-Vollhardt self-consistent equation for the diffusion coefficient. See A. M. Garcia-Garcia, Phys. Rev. Lett. {\bf 100}, 076404 (2008).
\end{thebibliography}
\end{document}